\documentclass[prl,twocolumn,superscriptaddress,showpacs,floatfix]{revtex4}

\usepackage{graphicx}
\usepackage{SIunits}
\usepackage[normal]{subfigure}

\begin{document}

\title{Observation of non-Markovian dynamics of a single quantum dot in a micropillar cavity}
\author{K.~H.~Madsen}\email{khoe@fotonik.dtu.dk}
\affiliation{DTU Fotonik, Department of Photonics Engineering, Technical University of Denmark, \O rsteds Plads 343, DK-2800 Kgs. Lyngby, Denmark}
\author{S.~Ates}
\affiliation{DTU Fotonik, Department of Photonics Engineering, Technical University of Denmark, \O rsteds Plads 343, DK-2800 Kgs. Lyngby, Denmark}
\author{T.~Lund-Hansen}
\affiliation{DTU Fotonik, Department of Photonics Engineering, Technical University of Denmark, \O rsteds Plads 343, DK-2800 Kgs. Lyngby, Denmark}
\author{A.~L\"{o}ffler}
\affiliation{Technische Physik, Universit\"{a}t W\"{u}rzburg, Am Hubland, D-97074 W\"{u}rzburg, Germany}
\author{S.~Reitzenstein}
\affiliation{Technische Physik, Universit\"{a}t W\"{u}rzburg, Am Hubland, D-97074 W\"{u}rzburg, Germany}
\author{A.~Forchel}
\affiliation{Technische Physik, Universit\"{a}t W\"{u}rzburg, Am Hubland, D-97074 W\"{u}rzburg, Germany}
\author{P.~Lodahl}\email{pelo@fotonik.dtu.dk} \homepage{www.fotonik.dtu.dk/quantumphotonics}
\affiliation{DTU Fotonik, Department of Photonics Engineering, Technical University of Denmark, \O rsteds Plads 343, DK-2800 Kgs. Lyngby, Denmark}

\date{\today}

\begin{abstract}
We measure the detuning-dependent dynamics of a quasi-resonantly excited single quantum dot coupled to a micropillar cavity. The system is modeled with the dissipative Jaynes-Cummings model where all experimental parameters are determined by explicit measurements. We observe non-Markovian dynamics when the quantum dot is tuned into resonance with the cavity leading to a non-exponential decay in time. Excellent agreement between experiment and theory is observed with no free parameters providing the first quantitative description of an all-solid-state cavity QED system based on quantum dot emitters.
\end{abstract}


\maketitle
All-solid-state cavity quantum electrodynamics (CQED) systems based on quantum dots (QDs) in nanophotonic cavities provide a promising platform for practical implementations of quantum information protocols \cite{Fattal.PRL.2004}. Solid-state systems are of inherently manybody nature, and a number of unique properties distinguish QD-based QED systems from their atomic counterparts including the presence of multiexciton states \cite{Winger.PRL.2009} and deviations from the point-dipole emitter description \cite{Andersen.NatPhys.2010}. It was recently observed in the spectral domain that the cavity-QD coupling is surprisingly efficient even when the QD is detuned several cavity linewidths away from resonance \cite{Hennessy.Nature.2007,Ates.Nature.2009}, and a qualitative understanding of such behavior is obtained by including the influence of phonons \cite{Naesby.PRA.2008,Kaer.PRL.2010,Auffeves.PRB.2010} and charged excitons \cite{Winger.PRL.2009}. Furthermore, the ability to enter the coherent and strong coupling regime was demonstrated by recording detuning-dependent emission spectra \cite{Reithmaier2004}. This proves, that the QD-cavity coupling is so strong that there is 'memory' in the system, i.e. the population of the QD at one instant of time depends on its value at previous times, which leads to the creation of light-matter entanglement. These so-called non-Markovian effects of the photon-matter coupling have been theoretically addressed in the broad context of photonic crystals \cite{Vats.PRA.2002} and appear in general when the local density of optical states (LDOS) is varying rapidly in frequency relative to the QD linewidth \cite{Kristensen.Opt.2008}. Non-Markovian phonon-matter effects have been experimentally demonstrated for single QDs in the spectral domain \cite{Galland.PRL.2008}.

In order to unambiguously identify non-Markovian effects, dynamical measurements are required. The experimental signature of non-Markovian coupling of a two-level emitter to a radiation bath is the deviation of the spontaneous emission intensity from an exponential decay in time. Such measurements are very demanding since reliable determination of decay curves from a single QD is required. The added benefit compared to spectral measurements, however, is that single QD decay curves are insensitive to the collection efficiency of the radiated intensity, which, e.g., depends sensitively on detuning in the experiment. Therefore dynamical measurements are expected to be superior for quantitative experiments on CQED systems. In this Letter, we present systematic time-resolved photoluminescence (PL), photon statistics, and two-photon interference measurements on a single QD in a micropillar cavity. The time-resolved measurements reveal non-Markovian dynamics, which to our knowledge is the first experimental demonstration in solid-state CQED, thus complementing results on atomic systems \cite{Brune.PRL.1996}. By combining the results of the different independent measurements, we explicitly extract all relevant parameters characterizing the QED system and obtain very good agreement between experiment and theory for the detuning-dependent dynamics without introducing any adjustable parameters. Such a detailed comparison between experiment and theory is crucial in order to unambiguously prove the existence of the delicate non-Markovian effects. We identify the QD-cavity system to be in an intermediate coupling regime close to the onset of coherent coupling, where non-Markovian effects are significant, as opposed to the widely studied weak coupling regime, where enhanced exponential decays are observed \cite{Gerard.PRL.1998,Santori.Nature.2002}.

A two-level emitter is described by the state $|\Psi(t) \rangle=c_e(t)|e,\{0\}\rangle+\sum_\mu c_\mu (t) |g, \{ \mu \} \rangle$, where $|e,\{0\}\rangle$ corresponds to the emitter in the excited state and a continuum of vacuum states, and $|g, \{ \mu \} \rangle$ the emitter in the ground state and one photon created in the mode $\mu$. The dynamics of the emitter follows from \cite{Vats.PRA.2002}
\begin{eqnarray} \label{ce}
\dot{c}_e(t)=-\int_0^t dt' c_e(t')\int_0^\infty d \omega e^{i(\omega_0-\omega)(t-t')} \omega \rho(\mathbf{r_0},\omega,\mathbf{\hat{e}_p})
\end{eqnarray}
where $\omega_0$ is the emitter frequency and $\rho(\mathbf{r_0},\omega,\mathbf{\hat{e}_p})$ is the projected LDOS at the position $\mathbf{r_0}$, frequency $\omega$, and orientation of the emitter transition dipole moment $\mathbf{\hat{e}_p}$. In most cases the Wigner-Weisskopf approximation applies, where the LDOS frequency dependence is sufficiently weak that it can be dragged outside the integration in Eq. (\ref{ce}) giving $\dot{c}_e(t) = -\left( \Gamma_{rad}/2 \right)c_e(t)$, corresponding to no memory in the system and an exponential decay of the population in time. However, in high-Q cavities or photonic crystals the frequency variation of the LDOS is strong, and the Wigner-Weisskopf approximation may break down leading to non-Markovian dynamics that mathematically implies that the time integration in Eq. (\ref{ce}) must be maintained. The physical interpretation of non-Markovian effects is, that the photon emitted through spontaneous emission acts back on the emitter.

\begin{figure}[tb]
\includegraphics[width=83mm]{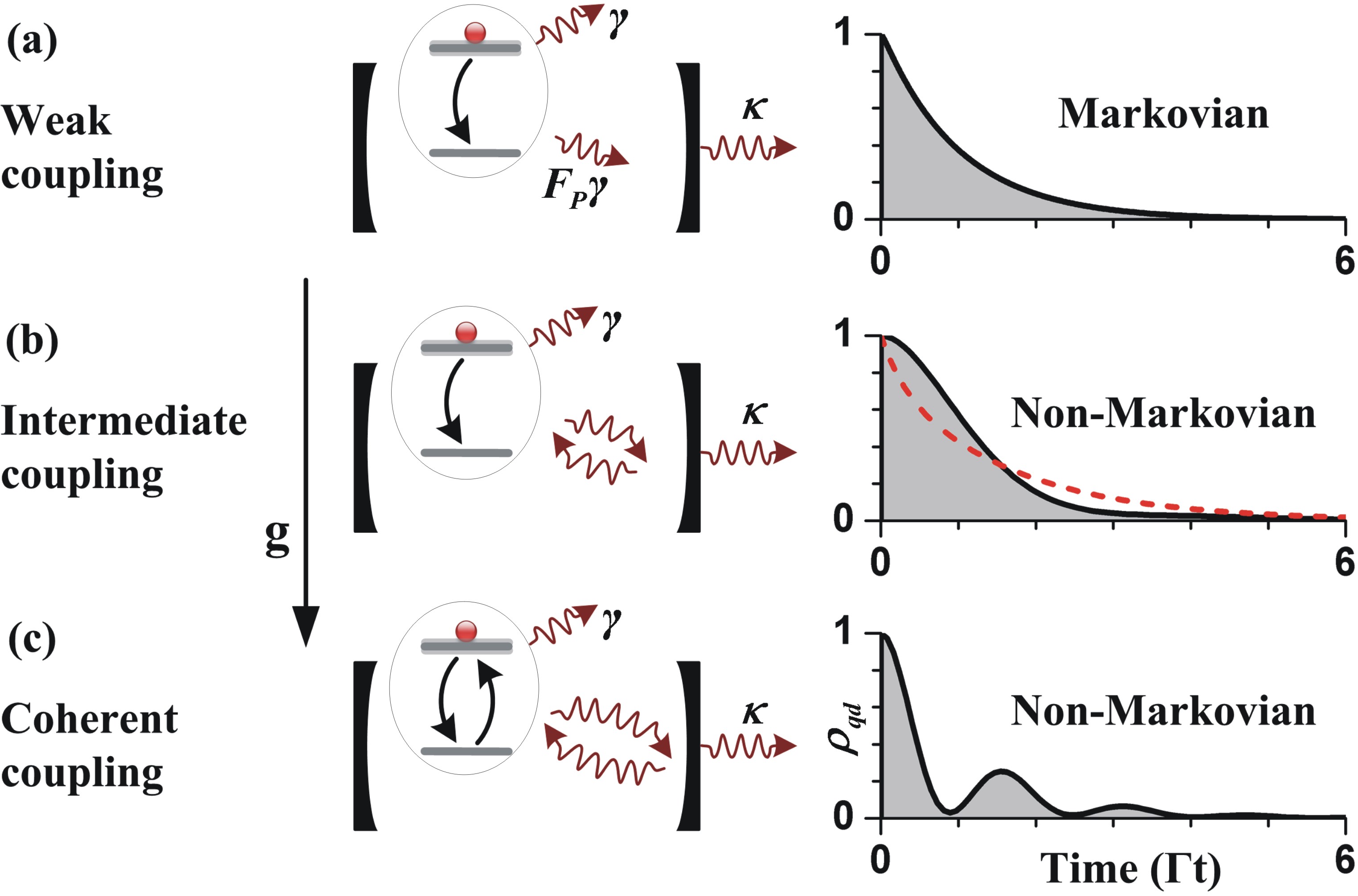}
\caption{(Color online). Illustration of the three distinct dynamical regimes for a coupled QD-cavity, with the physical system in the left panel and the resulting dynamics obtained from Eqs. (\ref{QDeq}) in the right panel. \textbf{a)} Weak-coupling regime: the QD decays exponentially with a Purcell enhanced rate ($F_P \gamma$). \textbf{b)} Intermediate regime: the QD dynamics is non-Markovian leading to a non-exponential decay in time. For reference the dashed curve shows an exponential decay. \textbf{c)} Coherent regime: the QD population undergoes Rabi oscillations. In regimes b) and c) the coherent back action of the radiation field on the QD leads to non-Markovian dynamics. \label{fig:sketch}}
\end{figure}

The special case of a Lorentzian LDOS in Eq. (\ref{ce}) corresponds to a single-mode cavity and the coupled QD-cavity system is described by the dissipative Jaynes-Cummings (J-C) model \cite{Carmichael1989}
\begin{eqnarray} \label{QDeq}
\dot{\rho}_{qd} &=& -\gamma \rho_{qd} - g^2 \left[ G(t)+G^*(t) \right] \\ \nonumber
\dot{\rho}_{ca} &=& -\kappa \rho_{ca} + g^2 \left[ G(t)+G^*(t) \right]\\ \nonumber
G(t) &=& \int^t_0 \left[ \rho_{qd}(t') -\rho_{ca}(t') \right] e^{-D(t-t')} dt'
\end{eqnarray}
where $D=\frac{\kappa+\gamma}{2}+\gamma_{dp}+i\Delta$, and $\rho_{qd}$ and $\rho_{ca}$ are the QD and cavity populations, respectively.
Furthermore, the cavity decay rate is related to the cavity Q factor through $\kappa=\omega_{ca}/Q$, $\gamma$ is the QD decay rate due to spontaneous emission out of the cavity, $g$ is the coupling strength, $\Delta=\omega_{ca}-\omega_{qd}$ is the cavity-QD frequency detuning, and $\gamma_{dp}$ is the phonon dephasing rate. This is an appropriate model for weak excitation, low temperatures, and a moderate $g$ \cite{Kaer.PRL.2010}. $G(t)$ is the memory kernel of the system. Three distinct dynamical regimes are identified cf. Fig. \ref{fig:sketch}: a) $|D|\gg 2g$ is the weak coupling Markovian regime where the QD decays exponentially in time with a Purcell enhanced rate. In this case there is no coherent back action of the cavity on the QD. b) $2g\lesssim |D|$ defines an intermediate regime where the Markovian Wigner-Weisskopf approximation breaks down. The resulting QD decay is irreversible but non-exponential due to the feedback from the cavity. c) $2g > |D|$ is the coherent and non-Markovian regime where the QD population undergoes vacuum Rabi oscillations due to the reexcitation of the QD resulting in a reversible evolution. The reversed arrows in Fig. \ref{fig:sketch} indicate the reversible dynamics and the influence of coherent back action from the cavity.

We have investigated a $1.7$ $\mu$m diameter micropillar cavity consisting of alternating GaAs and AlAs layers surrounding a central GaAs cavity and an embedded layer of low density InAs QDs ($60$-$90$ $\mu$m$^{-2}$) \cite{Loffler.APL.2005}. The sample is optically excited in a He flow cryostat under a $15$ degrees angle of incidence relative to the substrate surface using $~3$ ps long pulses from a Ti:Sapph. laser. The PL is collected with a microscope objective (N.A. = 0.6), spectrally resolved, and directed to either a CCD-camera, an APD, a Hanbury-Brown and Twiss (HBT) detector, or a Hong-Ou-Mandel (HOM) interferometer.

\begin{figure}[tp]
\includegraphics[scale=1]{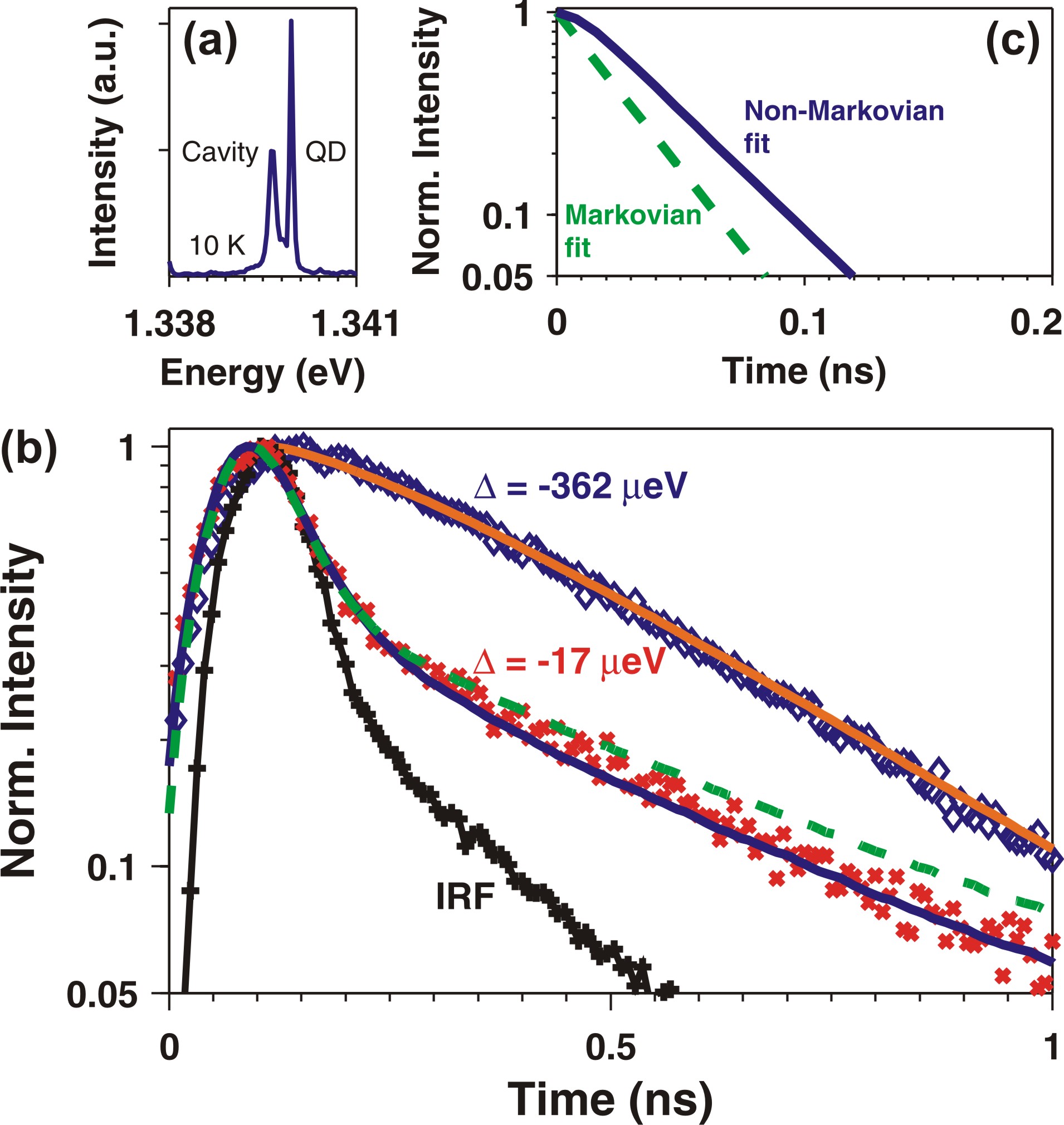}
\caption{(Color online). \textbf{a)} An example of an emission spectrum obtained with p-shell excitation at $T=10$ K. \textbf{b)} Measured decay curves close to and far from cavity resonance. The data recorded at $\Delta=-17$ $\mu$eV are modeled with the non-Markovian model obtained by solving Eqs. (\ref{QDeq}) (blue full line) with an additional slow exponential decay accounting for recombination of dark excitons. The green dashed curve represents the case where the Markov approximation has been assumed. The black curve shows the instrument response function (IRF) of the measurements that is used to convolute the theory in order to model the experimental data. Off resonance the Markov approximation is excellent (orange full line). \textbf{c)} The fast parts of the modeled decay curves recorded at $\Delta=-17$ $\mu$eV (blue full line) displaying a pronounced offset from the Markovian theory (green dotted curve).\label{fig:meas_decay_curves}}
\end{figure}

A typical emission spectrum of a QD coupled to the cavity under quasi-resonant (p-shell) pulsed excitation is shown in Fig. \ref{fig:meas_decay_curves}a, where the emission from a single QD together with the detuned fundamental cavity mode emission is clearly seen. Only emission from a single QD is observed due to the selective p-shell excitation, which is verified by autocorrelation measurements. The pronounced cavity emission observed even when the QD is off resonance has been reported previously~\cite{Kaniber2008,Press.PRL.2007}. The quality of the cavity is measured under strong above band excitation power to ensure all QDs are saturated \cite{Gayral2008}. By deconvoluting the spectrum with the spectrometer resolution, we find Q$=12200\pm410$ corresponding to $\kappa=166.7$ ns$^{-1}$. The QD dynamics was investigated by time-resolved PL measurements under systematic variation of the detuning ($\Delta$) via temperature tuning. Fig. \ref{fig:meas_decay_curves}b compares the decay of the QD for large detuning ($\Delta=-362$ $\mu$eV) and close to resonance ($\Delta=-17$ $\mu$eV). A large decrease in the lifetime is observed in the latter case due to the Purcell effect. The off-resonance decay curve is bi-exponential where the fast $(\gamma_f)$ and slow $(\gamma_s)$ decay rates correspond to the recombination of bright and dark excitons, respectively. We extract the radiative rate of spontaneous emission leakage out of the cavity as $\gamma = \gamma_f - \gamma_s =1.94$ ns$^{-1}$\cite{Johansen2010}.

\begin{figure}[t]
\includegraphics[scale=1]{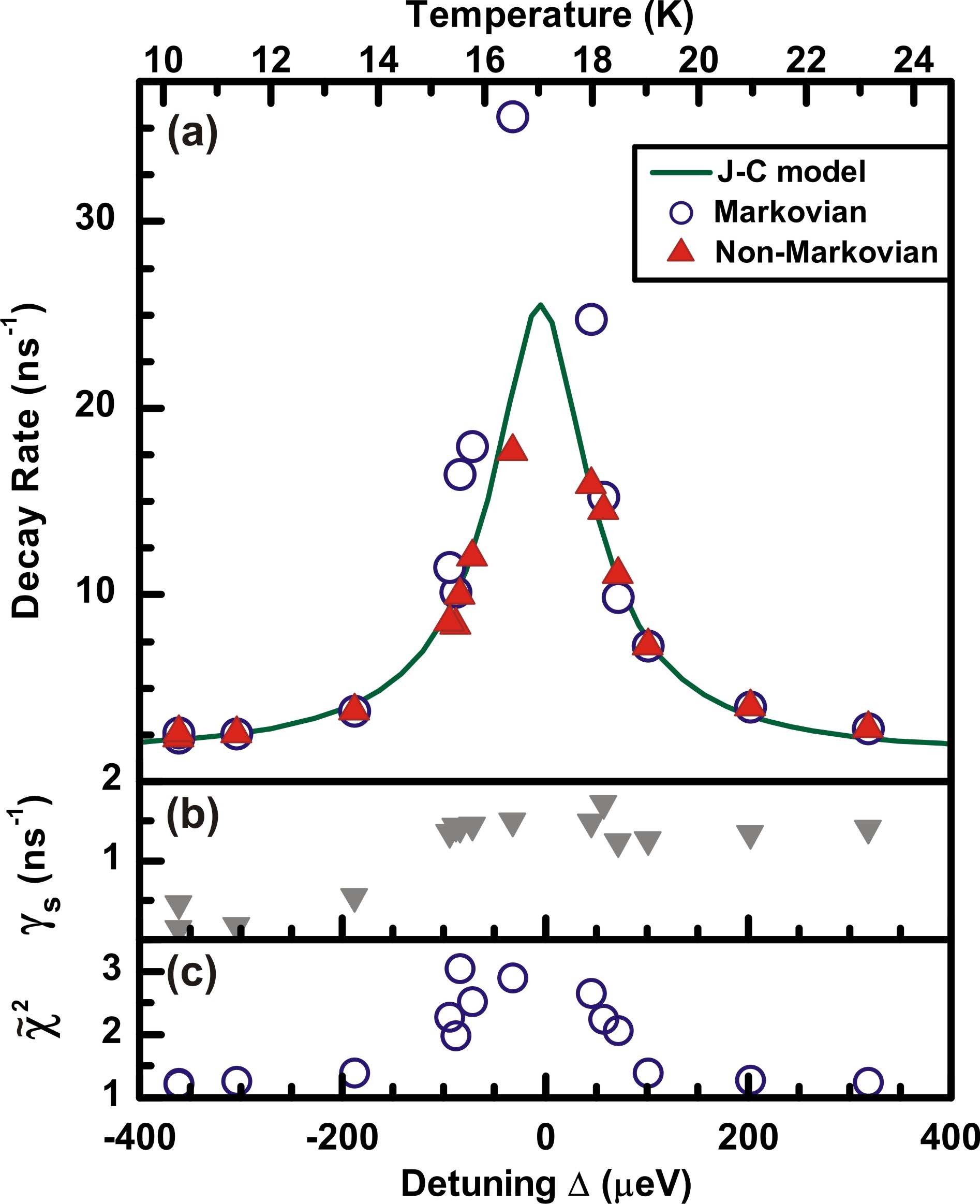}
\caption{(Color online). \textbf{a)} The detuning-dependent decay rate of the QD extracted by assuming the Markov approximation and with no approximations (non-Markovian). The large deviation between the two approaches clearly illustrates the importance of non-Markovian effects. The theory curve is calculated using only the experimentally measured parameters. \textbf{b)} The slow decay rate $\gamma_s$ and \textbf{c)} the reduced-chi-squared $\tilde{\chi}^2$ for the Markovian fits versus detuning obtained by modeling decay curves like the ones shown in Fig. \ref{fig:meas_decay_curves}b.\label{fig:mean_rates}}
\end{figure}

Close to resonance non-Markovian effects are observed. We stress that the non-Markovian effects are rather challenging to observe since even for an infinitely fast detector it would be very difficult to observe the variations in the slope between the Markovian and non-Markovian results, see Fig. \ref{fig:meas_decay_curves}c. However, a clear offset between the two curves would appear and this constitutes the tell-tale of non-Markovian effects in the intermediate coupling regime. In order to reliably prove the presence of such an offset, it is imperative to measure all experimental parameters for the system in order to have a detailed comparison between experiment and theory, which is exactly the approach of the present work. Close to resonance, the decay curves are modeled with the solution obtained from Eq. (\ref{QDeq}), where the experimental values of $\gamma, \kappa$, and $\gamma_{dp}$ (see below) are used. Additionally, a slow component with decay rate $\gamma_s$ accounts for dark exciton recombination, and the theory is convoluted with the measured IRF. Fig. \ref{fig:meas_decay_curves}b (blue curve) shows the comparison between experiment and theory close to resonance ($\Delta=-17$ $\mu$eV), from which we extract the coupling strength $g=34.3 \pm 1.4$ ns$^{-1}$. We evaluate $|D| = 97.4$ ns$^{-1}$, which is in the intermediate coupling regime $(2g\lesssim |D|)$. Describing the data using the Markovian bi-exponential model is shown in Fig. \ref{fig:meas_decay_curves}b (dotted green curve), where the same slow component $\gamma_s$ is used. Erroneously assuming the Markovian approximation to hold thus implies a significant deviation between theory and experiment that is clearly visible in the raw decay curve (see Fig. \ref{fig:meas_decay_curves}b) indeed giving rise to the offset discussed above. The model decay curves plotted without the contributions of the IRF of the measurement deviate even more, see Fig. \ref{fig:meas_decay_curves}c. The inability to model the decay curves with a Markovian theory is quantified by the goodness-of-the-fit displayed in Fig. \ref{fig:mean_rates}c where significant deviations from the optimum value of unity is observed close to resonance.

An alternative way to display the deviations from Markovian behavior observed in the experiment is revealed by comparing the characteristic QD decay rates to theory. In the non-Markovian regime the decay curves are multi-exponential, and we extract the mean decay rate ${1 / \left< t \right>} = \int_0^\infty \rho_{qd} dt / \int_0^\infty t \cdot \rho_{qd} dt$ after excluding the slow components of the decay curves that are due to dark excitons. We note that the slow component is modeled for each decay curve individually, and in Fig. \ref{fig:mean_rates}b a weak temperature dependence is observed. Fig. \ref{fig:mean_rates}a displays a very pronounced deviation close to resonance when comparing the rates extracted with and without the Markov approximation. Importantly the rates extracted with the full non-Markovian model of the decay curves are found to be in excellent agreement with the predictions from the J-C model without assuming any adjustable parameters, as explained below. Mistakenly assuming the Markov approximation to hold leads to large over estimations of the decay rate being $35.6$ ns$^{-1}$ at resonance while the correct value of the mean decay rate in the full non-Markovian model is $17.7$ ns$^{-1}$. Thus the failure to account for non-Markovian effects would in fact significantly influence the extracted value of the coupling strength $g$ leading to the incorrect conclusion that the cavity would be even closer to the coherent coupling regime where non-Markovian effects would in fact be even more pronounced.
The comparison to theory has been undertaken by calculating the detuning-dependent dynamics of the QD with the J-C model of Eq. (\ref{QDeq}) using all the independently measured parameters, see Fig. \ref{fig:mean_rates}a. An excellent agreement between experiment and theory is apparent where we stress that no free parameters have been introduced, making this the first quantitative understanding of a QD-based CQED system. Previous comparisons between experiment and theory have not extracted the parameters independently, but rather employed global-fitting routines \cite{Laucht.PRL.2009} or assumed reasonable values for the relevant parameters \cite{Hohenester.PRB.2009}. In our experiment, the full detuning-dependence of the dynamics is explained by the J-C model with dissipation and dephasing. We record the full-width-half-maximum of the regime with Purcell enhancement to be $99.3$ $\mu$eV in Fig. \ref{fig:mean_rates}a, which is larger than the value $88.3$ $\mu$eV obtained from a model without dephasing. This is the experimental proof of the broadening of the Purcell effect due to pure dephasing, as was predicted in \cite{Auffeves.PRB.2010}.

Finally we discuss how the remaining experimental parameters required for the comparison between experiment and theory were extracted. The single-emitter nature of the QD emission was verified by pulsed autocorrelation measurements using an HBT setup. For large detuning we obtain $g_{\Delta \gg 0}^{(2)}(0)$=$13.2\%$, which quantifies the contribution of multi-photon emission on the identified emission line. The observed excess photons could originate from residual excitation of different charge configurations in the QD that have been found to be a pronounced effect with non-resonant excitation schemes \cite{Winger.PRL.2009}.The coherence of the emitted single photons were measured with an HOM interferometer~\cite{Santori.Nature.2002}, where the coincidence counts at zero electronic delay between two single-photon counters is related to the wavepacket overlap (or visibility, $V$) of two consecutively emitted photons. We extract $V=48\%$. From $V$ we can extract the pure dephasing rate of the QD \cite{Bylander2003}, and we obtain $\gamma_{dp}=(6.3 \pm 2.2)$ $\mu$eV at $T=16.3$ K. Below $60$ K, the dephasing rate has been shown to depend linearly on temperature~\cite{Borri2001}, and we therefore have $\partial \gamma_{dp}/\partial T=(0.39 \pm 0.13)$ $\mu$eV/K, that is valid in the temperature range of the experiment (10-23 K). We note that only the temperature dependence of the dephasing rate needs to be considered, since all other parameters vary very little in the applied temperature range, which is apparent from the excellent agreement between experiment and theory.

In conclusion, we have observed non-Markovian dynamics of a QD coupled to a micropillar cavity. The non-Markovian dynamics is observed close to cavity resonance, and the experimental signature is multi-exponential decay dynamics leading to an observable offset in the decay curves compared to the Markovian result. Failing to account for non-Markovian effects were found to lead to a large overestimate of the coupling strength. Finally, we observed excellent agreement between theory and experiment for the detuning-dependent decay rates without assuming any free parameters, thus providing the first quantitative understanding of a QD-based CQED system.

We gratefully acknowledge financial support from the Villum Kann Rasmussen Foundation and The Danish Council for Independent Research (Natural Sciences and Technology and Production Sciences). This work was part of the EU project "QPhoton". M. Emmerling and A. Wolf are acknowledged for sample preparation.

\end{document}